\documentclass[prb,superscriptaddress,showpacs,showkeys,twocolumn,10pt]{revtex4-1}
\usepackage{amsfonts,amsmath,amssymb,graphicx,epstopdf,verbatim,color}
\usepackage[english]{babel}

\newcommand{\kk}{k}

\begin{document}

\title{Dynamical quantum depletion in polariton condensates}
\author{Selma Koghee}
\affiliation{Theory of Quantum and Complex Systems, Universiteit Antwerpen, Universiteitsplein 1, B-2610 Antwerpen, Belgium}
\author{Michiel Wouters}
\affiliation{Theory of Quantum and Complex Systems, Universiteit Antwerpen, Universiteitsplein 1, B-2610 Antwerpen, Belgium}
\date{\today}

\begin{abstract}

We present a theoretical study of the quantum depletion of microcavity polaritons that are excited with a resonant laser pulse. The dynamics of the quantum fluctuations are interpreted in the context of quantum quenches in general and in terms of the dynamical Casimir effect in particular. We compute the time evolution of the first and second order correlation functions of the polariton condensate. 
Our theoretical modelling is based on the truncated Wigner approximation for interacting Bose gases. For homogeneous systems, analytical results are obtained in the linearised Bogoliubov approximation. Inhomogeneous systems are studied numerically by Monte Carlo simulations.
\end{abstract}

\maketitle

\section{Introduction}

Interaction quenches in quantum many body systems have become an active research field \cite{polkovnikovRMP,schmiedmayer14}, mainly thanks to the great degree of controllability of ultracold atoms with Feshbach resonances and optical lattices \cite{BDZ}.
Recently, a complementary platform for quantum many body physics that has been developed, namely exciton-polariton quantum fluids \cite{iacrev}. The distinctive characteristic of these systems is that the polaritons are a superposition of light and matter excitations. A first advantage of the light component is that it allows for a straightforward diagnostic of the fluid by means of standard quantum optical techniques. A second advantage is that polaritons can be created by an external laser field. It is this feature of polariton condensates that is of particular interest in the context of quantum quenches, since it allows to implement a sudden change in the many body system. 

The situation that we will consider here is an instantaneous injection of polaritons in a coherent state \cite{koghee14}. Since this is not the ground state of the interacting many-body system, a non-trivial time evolution will result. In our previous work, we have shown that a dynamical Casimir effect \cite{more70,dodonov,wilson11} takes place in terms of the Bogoliubov excitations on top of the coherently created polariton state. Indeed, the sudden creation of a condensate quenches the vacuum from the trivial one to the Bogoliubov vacuum, resulting in an excitation of the system. Part of the motivation for the study of the dynamical Casimir effect stems from connections with the Hawking-Unruh effect, whose sonic version \cite{unruh81} is getting within the reach of experiments with polaritons \cite{nguyen15} and ultracold atoms \cite{steinhauer14}.

The analogy with an interaction quench in cold atom systems is direct, since our proposal is equivalent to a sudden increase of the interaction strength, from zero to a finite value. Such experiments have been performed with ultracold atoms, for example,  Hung \textit{et al.} \cite{chin12} suddenly decreased the interaction strength in a weakly interacting atomic Bose-Einstein condensate . The resulting density oscillations were related to Sakharov oscillations in the early universe \cite{sakharov}. 

An even closer connection can be made with the splitting quench by Langen {\em et al.} in one-dimensional atomic condensates \cite{langen}. When a condensate is rapidly split in two parts, there is initially perfect phase coherence between them. However, at later times, the two parts start to develop a different phase. This dephasing due to interactions is entirely analogous to the one in our dynamical Casimir proposal, showing a light-cone-like emergence of thermal correlations.

An important difference between polaritonic and atomic condensates concerns the ratio of the life time with respect to the characteristic time scale of the dynamics. Whereas for ultracold atoms, this ratio is very large, in polariton systems losses are more important. Their theoretical modelling should therefore be carried out in an open system setting. This raises the interesting issue of the competition between losses and thermalization dynamics.

We will treat the open system quantum dynamics within the truncated Wigner approximation, which is a popular tool in both the study of conservative cold atoms \cite{sinatra02,polkovnikov10} as for lossy polariton systems \cite{iacrev}. 
When the condensate depletion is small, the equations of motion can be linearized in the fluctuations, which is equivalent to the Bogoliubov approximation. 

In Sec. \ref{sec:hom}, we use this approximation allows to obtain analytical results for  the first and second order coherence functions in the homogeneous system. For the inhomogeneous case instead, we perform in Sec. \ref{sec:inhom} Monte Carlo simulations of the stochastic equations of motion. We show that for a large smooth pumping spot, local density approximation satisfactorily reproduces the first order coherence function. Conclusions are drawn in Sec. \ref{sec:concl}.

\section{The Model \label{sec:hom}}

When a microcavity is excited sufficiently close to the lower polariton branch and all the relevant energy scales (linewidth, interaction energy) are much smaller than the Rabi splitting, it is well justified to restrict the dynamics to the lower polariton branch.

We consider a driven dissipative bosonic system, whose dynamics is governed by a master equation of the Lindblad type
\begin{equation}
\frac{d}{dt} \rho =  -\frac{i}{\hbar} [H, \rho] + \mathcal D(\rho).
\label{eq:master}
\end{equation}
Here, the Hamiltonian $H=H_{P} + H_{L}$  contains the free Bose gas dynamics of the polaritons
\begin{equation}
H_{P} = \int dx \; \psi^\dag(x) \left[ \frac{-\hbar^2}{2m} \nabla^2
+ \frac{g}{2} \psi^\dag(x) \psi(x) \right]  \psi(x),
\end{equation}
with $m$ being the lower polariton effective mass and $g$ the interaction strength, and it includes the external laser driving 
\begin{equation}
H_L=\int dx \; \left[ F_L(x,t) \psi^\dag(x) + F^*_L(x,t) \psi(x) \right],
\end{equation}
where $F_L$ is the laser amplitude.

The polariton losses, which depend on the linewidth $\gamma$, are described by the dissipator $\mathcal D(\rho)$, that we take to be of Lindblad form
\begin{eqnarray}
\mathcal D(\rho) &=& \int dx \;\frac{\gamma}{2 \hbar} \big[ 2 \psi(x) \rho \psi^\dag(x) \nonumber \\ 
& &- \psi^\dag(x) \psi(x) \rho -  \rho \psi^\dag(x) \psi(x) \big].
\end{eqnarray}

For the quantum quench that we consider, we take the driving laser to be an ultra short pulse. When the pulse duration $\delta t$ is much shorter than all the other time scales of the dynamics, shortly after the pulse, the polariton field is in a coherent state with amplitude
$\psi_0 \equiv \langle \psi(x,t=0) \rangle  = \int_{-\delta t}^0 F_L(x,t) dt$. The external laser drive then only sets the initial condition and does not affect the polariton dynamics, which is governed by the free Bose gas dynamics and the losses only. 

We will solve the master equation \eqref{eq:master} within the truncated Wigner approximation (TWA), a method that is widely used for the simulation of weakly interacting one-dimensional atomic condensates. The addition of losses makes the TWA even a better approximation to the exact dynamics.

The resulting stochastic equations of motion read \cite{koghee14}:
\begin{align}
i\hbar \, d \phi(x,t) =& \left[-\frac{ \hbar^2 \nabla^2}{2m} - i \frac{\gamma}{2} + g |\phi(x,t)|^2 \right] \phi(x,t) \, dt \nonumber \\
&+ \sqrt{\frac{\hbar \gamma}{4 \Delta V}} dW(x,t),
\label{eq:wigev}
\end{align}
where $\Delta V$ is the volume of a single cell of the discretized grid. Since the expectation values of the stochastic fields are equal to the symmetrised averages of the quantum fields, the TWA can be used to study the quantum fields.

\section{Homogeneous system}

\subsection{Bogoliubov approximation}

As long as the condensate depletion is small, the dynamics can be treated in the linearised Bogoliubov approximation. The field $\phi(x,t)$ is decomposed in Fourier space 
\begin{equation}
\phi(x,t)=\phi_c(t) + \frac{1}{\sqrt{L}}\sum_k \phi(k,t) e^{i k x},
\end{equation}
where $L$ is the length of the one dimensional wire that we consider. The evolution of the condensate density is determined to be
\begin{equation}
\langle \phi_c^*(t) \phi_c(t) \rangle \equiv n_c(t) = n_c(0) \exp\left( - \gamma t \hbar \right)
\end{equation}
and the equations of motion for the fluctuations are linearised in $\phi(k,t)$.
In terms of the vector $\Phi(k) = [\phi(k),\; \phi(-k)]^T $ and the noise vector $d \Xi(k) = [dW(k),\; dW(-k)]^T$, they read
\begin{equation}
i \hbar \, d \Phi(k) = B(k,t) \Phi(k) dt + \frac{\sqrt{\hbar\,\gamma}}{2} d\Xi(k),
\label{eq:bogdyn}
\end{equation}
where the Bogoliubov matrix equals
\begin{eqnarray}
\!\! B(k,t) = \!
\begin{pmatrix}
\epsilon(k) + gn_c(t) -\frac{i \gamma}{2} & g n_c(t) \\
-g n_c(t) & -\epsilon(k) - gn_c(t) -\frac{i \gamma}{2}
\end{pmatrix} 
\label{eq:bogmat}
\end{eqnarray}
and $\epsilon(k) = \hbar^2 k^2 / (2m)$. From the solution of these stochastic differential equations, we can compute the time evolution of the correlation functions.

\subsection{Momentum distribution and first order coherence}

The differential equation \eqref{eq:bogdyn} for the stochastic fields can be solved exactly in the limit $k \rightarrow 0$, which gives for the momentum distribution
\begin{equation}
\lim_{k\rightarrow 0} n(k,t) = 2 \left(\frac{g n_c(0)}{\gamma}
\right)^2 e^{-2\gamma t/\hbar} \left(e^{\gamma t/\hbar} -\frac{\gamma t}{\hbar} -1 \right).
\label{eq:NK0}
\end{equation}

For large momenta, we will resort to the sudden approximation \cite{carusotto10}, which has yielded a good description of the average value of the momentum distribution:
\begin{eqnarray}
&&\langle \psi^\dagger(k,t)\psi(k,t)\rangle = \nonumber \\
&&\left[ \frac{g n_c(0)]}{\hbar \omega_B(k)} \right]^2 \sin\left[ \hbar \omega_B(k) t \right]^2 e^{-\gamma t/ \hbar},
\label{eq:momdis}
\end{eqnarray}
where $\hbar \omega_B(k)=\sqrt{\epsilon(k)\,[\epsilon(k)+2g n_c(0)]}$ is the Bogoliubov dispersion. In the next section, we will calculate the second order coherence from the momentum distribution and the anomalous average  $\langle \psi(k,t)\psi(-k,t)\rangle$. The latter quantity is calculated following the same procedure as for the momentum distribution. Thus, we first determine $\langle \psi(k,t)\psi(-k,t)\rangle$ for a system without decay, i.e. $\gamma =0$. and reintroduce the time dependence by letting the expectation values decay exponentially. This yields
\begin{eqnarray}
& \langle \psi(k,t)\psi(-k,t)\rangle = 
-  \frac{ g n_c(0) \sin\left[ \hbar \omega_B(k) t \right]}{\left[ \hbar \omega_B(k) \right]^2} e^{-\gamma t/ \hbar}  & \label{eq:anomav} \\
& \left\{ \left[ \epsilon(k) + g n_c(0) \right] \sin\left[ \hbar \omega_B(k) t \right] 
+ i \hbar \omega_B(k) \cos\left[ \hbar \omega_B(k) t \right] \right\}. &\nonumber
\end{eqnarray}

The Fourier transform of the momentum distribution gives us the first order correlation in real space
\begin{equation}
g^{(1)}(x,x')= \frac{\langle \psi^\dag(x,t) \psi(x',t) \rangle}{\sqrt{\langle \psi^\dag(x,t) \psi(x,t) \rangle  \langle \psi^\dag(x',t) \psi(x',t) \rangle}}.
\end{equation}
From this quantity, we can obtain the condensate fraction, the quantum depletion $\delta n/n_c$, 
\begin{equation}
\frac{\delta n(t)}{n_c(t)}
= C \; \frac{ g^2 n_c(0)}{\gamma^2} \left[1- e^{-\gamma t/\hbar} \left(\frac{\gamma t}{\hbar} +1 \right) \right] k_*(t),
\label{eq:dnscale}
\end{equation}
with 
\begin{equation}
k_*(t)=\frac{\gamma}{2 \hbar} \sqrt{\frac{m}{g n_c(0)}} \left[1- e^{-\gamma t/\hbar} \left(\frac{\gamma t}{\hbar} +1 \right) \right]^{-1/2},
\label{eq:kstar}
\end{equation}
and the coherence length $\ell_c$
\begin{equation}
\ell_c(t) = 2.1 / k_*(t),
\label{eq:cohlength}
\end{equation}
where the factor 2.1 was determined numerically.

\subsection{Second order coherence in momentum space}
Since the particles are predicted to be produced in pairs with opposite momentum, we expect to find a correlation between polaritons with momentum $k$ and those with momentum $-k$. Therefore, we will study the second order coherence in momentum space:
\begin{eqnarray}
&& g^{(2)}(k,-k,t) = 
\label{eq:g2basic} \\
&& \frac{ \langle \psi^\dagger(k,t) \psi^\dagger(-k,t) \psi(-k,t) \psi(k,t) \rangle }{ \langle \psi^\dagger(k,t) \psi(k,t) \rangle \langle \psi^\dagger(-k,t) \psi(-k,t) \rangle }. \nonumber
\end{eqnarray}
By applying Wick contraction, we can write the denominator as a product of quadratic expectation values. The nonzero terms are those containing the momentum distribution $\langle \psi^\dagger(k,t) \psi(k,t) \rangle$ and the anomalous average $\langle \psi^\dagger(k,t) \psi^\dagger(-k,t) \rangle$. In terms of the stochastic fields, the expression becomes
\begin{eqnarray}
&&\langle \psi^\dagger(k,t) \psi^\dagger(-k,t) \psi(-k,t) \psi(k,t) \rangle = \nonumber \\
&&\langle \phi^*(k,t) \phi(k,t) \rangle \langle \phi^*(-k,t) \phi(-k,t) \rangle 
\label{eq:fouropwick} \\
+&&\langle \phi^*(k,t) \phi^*(-k,t) \rangle \langle \phi(-k,t) \phi(k,t) \rangle \nonumber \\
&& -\frac{1}{2} \langle \phi^*(k,t) \phi(k,t) \rangle -\frac{1}{2} \langle \phi^*(-k,t) \phi(-k,t) \rangle + \frac{1}{4}. \nonumber 
\end{eqnarray}
In the limit $k \rightarrow 0$, an exact solution of equation \eqref{eq:bogdyn} can be found, which yields for the second order coherence

\begin{widetext}
\begin{eqnarray}
&& \lim_{k \rightarrow 0} g^{(2)}(k,-k,t) = 2 +\\
&& \frac{(\gamma t / \hbar)^2 \exp \left( -2 \gamma t / \hbar \right) }{4 (g n_c(0) / \gamma)^2 \left[ (1+ \gamma t / \hbar)^2 \exp \left( -4 \gamma t / \hbar \right) - 2(1+ \gamma t / \hbar) \exp \left( -3 \gamma t / \hbar \right) +\exp \left( -2 \gamma t / \hbar \right) \right]}. \nonumber
\end{eqnarray}
\end{widetext}

The fraction in this expression diverges both at short times ($t \ll \hbar/\gamma$), when $\lim_{k \rightarrow 0} g^{(2)}(k,-k,t) \approx 2 + (g n_c(0) t/\hbar)^{-2}$ and for long times ($t \gg \hbar/\gamma$), when $\lim_{k \rightarrow 0} g^{(2)}(k,-k,t) \approx 2 + (\gamma^2 t/\hbar g n_c(0))^2$. However, for good cavities with $\gamma \ll g n_c(0)$, there is a large time window when $\lim_{k \rightarrow 0} g^{(2)}(k,-k,t) \approx 2$. 

For large momenta, an expression for the second order coherence can be calculated from equation \eqref{eq:fouropwick} and the results obtained with the sudden approximation \eqref{eq:momdis} and \eqref{eq:anomav}. Combining these solutions and averaging out the oscillations afterwards, yields the following expression for the second order coherence
\begin{equation}
g^{(2)}(k,-k,t) = 2 + 2 \left[ \frac{ \hbar \omega_B(k)}{ g n_c(0)]} \right]^2.
\label{eq:g2analyticexpr}
\end{equation}
Here, we recognise the inverse of the momentum distribution without the exponential decay. This analytical result can be well understood from the assumption that the particles are indeed produced in pairs. In this case, a polariton with momentum $k$ will always be accompanied by a polariton with momentum $-k$. For large momenta, we expect very few particles. When the expected value is much smaller than one, only in few of the realisations a polariton will be present and the cases with more than one polariton with the same momentum will be negligible. Therefore, we find that when the polaritons are produced in pairs with opposite momentum, when  $\psi^\dagger(k,t)\psi(k,t) \ll 1$, and neglecting the exponential decay,
\begin{equation}
\langle \psi^\dagger(k,t) \psi^\dagger(-k,t) \psi(-k,t) \psi(k,t) \rangle =  \langle \psi^\dagger(k,t) \psi(k,t) \rangle .
\end{equation}
As a result, the second order coherence is the inverse of the momentum distribution
\begin{equation}
g^{(2)}(k,-k,t) \big|_{\textrm{large} \, k} = \frac{ 1 }{ \langle \psi^\dagger(k,t) \psi(k,t) \rangle },
\end{equation}
which is in close agreement with \eqref{eq:g2analyticexpr}.

\begin{figure}
	\includegraphics[scale=0.2, angle=0]{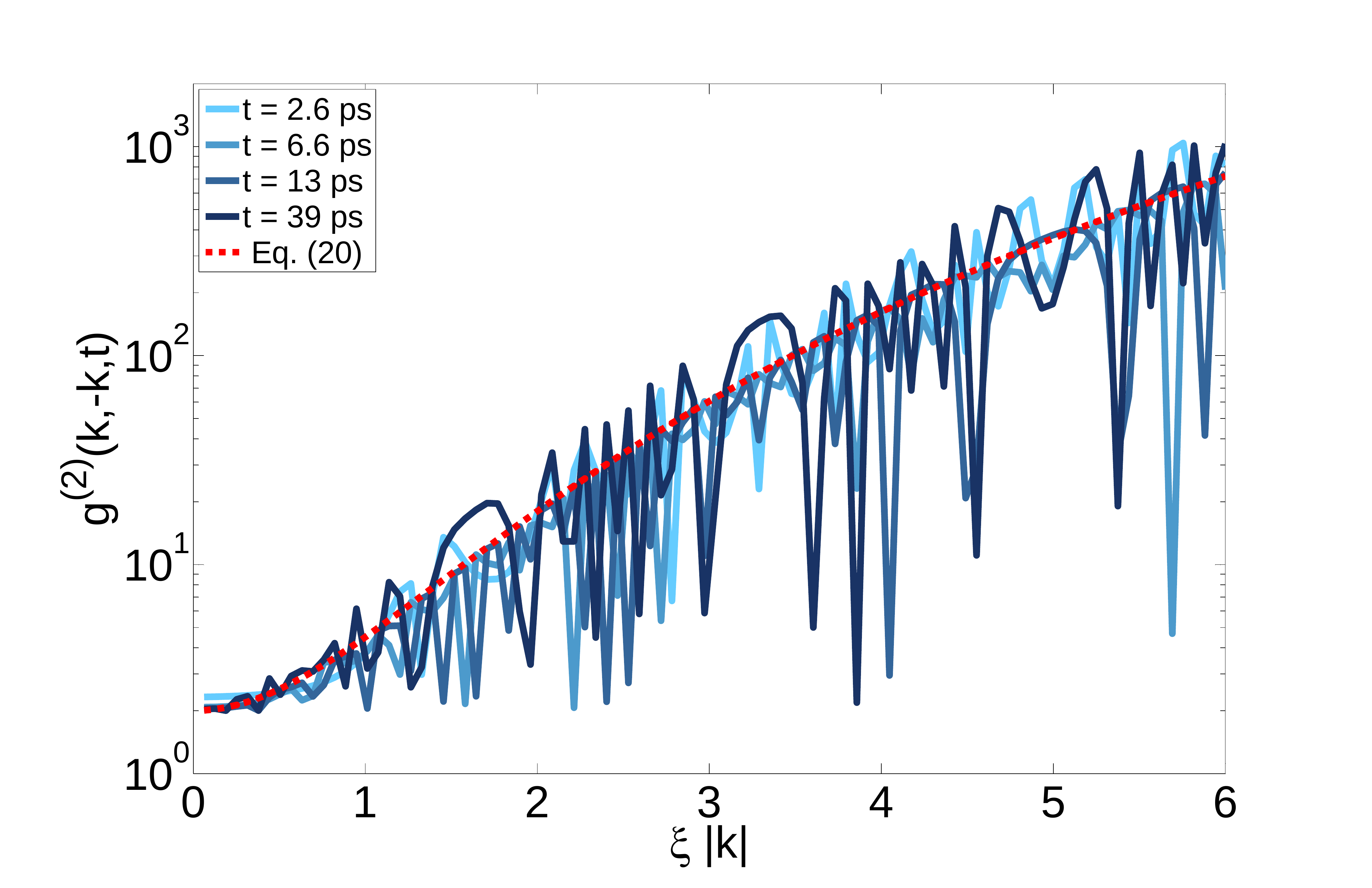}
	\caption{Second order coherence in momentum space for a homogeneous system calculated with the Green's function method. Solid lines show the numerical results, whereas the dotted line represents eq. \eqref{eq:g2analyticexpr}.
We have chosen $g=0.01 \, \mu {\rm m \, meV}, \; \gamma=0.05 \, {\rm  meV}, \hbar=1, \; m=1$, $g n_c(0)/\gamma=10$.		
	 }
	\label{fig:g2homGF}
\end{figure}

For the homogeneous system, the numerical calculations have been performed using the Green's function method from \cite{koghee14}. In terms of the Green's function 
\begin{equation}
G_\kk(t,t')=\prod_{j=1}^N \exp[-i \Delta t B_\kk(t_j)/\hbar],
\label{eq:greensfunction}
\end{equation} 
the second order coherence can be written 
\begin{eqnarray}
&&g^{(2)}(k,-k,t) = 1 +  \nonumber \\
&& \frac{1}{4} \bigg(
\left[ G^\dagger_\kk(t,0) \right]_{1,2} \left[ G_\kk(t,0) \right]_{1,1} +\left[ G^\dagger_\kk(t,0) \right]_{2,2} \left[ G_\kk(t,0) \right]_{1,2} \nonumber \\
&&+ \frac{\gamma}{\hbar} \int_{0}^t ds \bigg\{ 
\left[ G^\dagger_\kk(t,s) \right]_{1,2} \left[ G_\kk(t,s) \right]_{1,1} 
\label{eq:g2gf} \\
&&+\left[ G^\dagger_\kk(t,s) \right]_{2,2} \left[ G_\kk(t,s) \right]_{1,2} \bigg\} \bigg) \nonumber \\
&& \times \left[ \langle \psi^\dagger(k,t) \psi(k,t) \rangle \langle \psi^\dagger(-k,t) \psi(-k,t) \rangle \right]^{-1}, \nonumber
\end{eqnarray}
where the $x,y$ of $\left[ G_\kk(t,s) \right]_x,y$ indicate the matrix component. 
This results is depicted in figure \ref{fig:g2homGF}, together with the analytical expression eq. \eqref{eq:g2analyticexpr} derived from the sudden approximation. It can be seen that the analytical expression describes the overall behaviour of $g^{(2)}(k,-k,t)$ very well. The fast oscillations of the numerical results are expected to become more averaged in experimental data, which would then become closer to the analytically calculated average.

\section{Inhomogeneous sytem  \label{sec:inhom}}

\subsection{Monte Carlo simulation}

Although the Green's function method provides a good description of the homogeneous system, it has some limitations. First, the interactions energy should not be too high, since the Bogoliubov  approximation is no longer valid when the quantum depletion becomes too large. Secondly, the Green's function method becomes cumbersome for inhomogeneous systems. In order to overcome these problems, we have also implemented a Monte Carlo simulation algorithm \cite{carusotto05}. 

In the truncated Wigner Monte Carlo algorithm, the expectation values are calculated by averaging over many realisations of the system. The results presented here are obtained from 10000 realisations. For the initial situation, an average density is chosen and random noise is added to account for the stochastic nature of the fields. As opposed to the Green's function method, where equation $\eqref{eq:bogdyn}$ was solved, the Monte Carlo algorithm used both the real and momentum space representation of the stochastic fields. The evolution due to interactions and decay, which are the time-dependent parts of the Hamiltonian, is calculated in real space, whereas the effect of the kinetic term is calculated in momentum space. This method has the advantage that we do not have to distinguish between condensate and excitations in the interaction term. In order for the algorithm to work, the time steps for which the evolution is calculated should be small, in order for the effect of sequentially calculating the evolution in real an momentum space to be small. For the homogeneous system we have used a system length of $200 \, \mu\textrm{m}$ and $0.4 \, \mu\textrm{m}$ as the size of a unit cell. For the inhomogeneous systems, we adapted the length and grid size in order to have sufficient detail, the boundaries of the system distant enough with respect to the width of the Gaussian distribution, while keeping the number of grid points equal to 265 for computational efficiency. 

\begin{figure}
	\includegraphics[scale=0.2, angle=0]{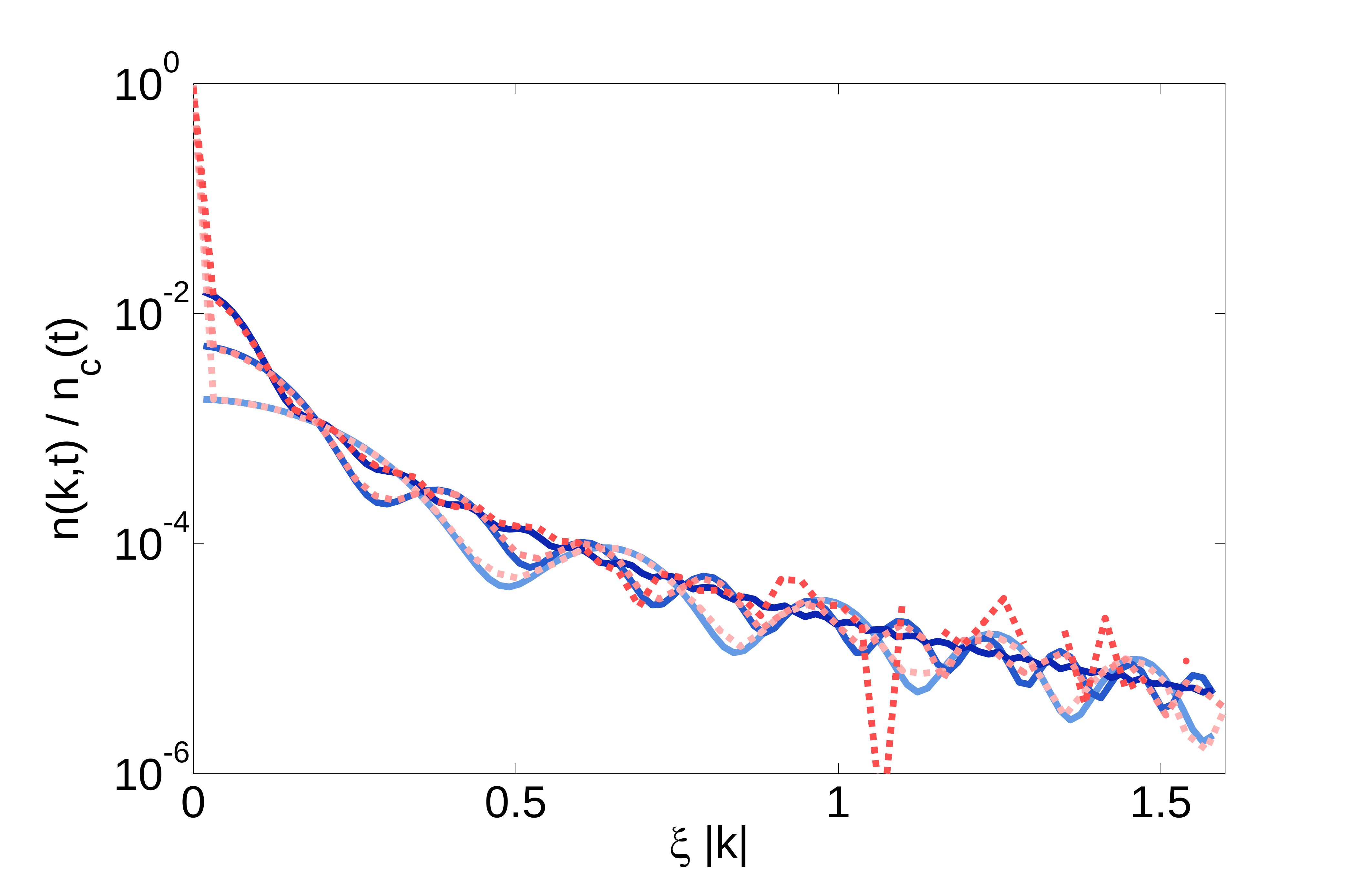}
	\caption{Momentum distribution for a homogeneous system calculated with the Greens' function method (blue solid) and Monte Carlo simulations (red dotted). We have chosen $g=0.01 \, \mu {\rm m \, meV}, \; \gamma=0.05 \, {\rm  meV}, L= 200 \, \mu {\rm m}, \hbar=1, \; m=1$, $g n_c(0)/\gamma=10$.}
	\label{fig:momdishomGFQMC}
\end{figure}

First, we verified the results of the Green's function method for the homogeneous system, see Fig. \ref{fig:momdishomGFQMC}. Secondly, we studied the effect of a Gaussian density distribution.
For the second order coherence, the expectation value of four operators can be computed directly using the Monte Carlo simulations as
\begin{eqnarray}
&&\langle \psi^{\dagger}(k,t) \psi^{\dagger}(-k,t) \psi(-k,t) \psi(k,t) \rangle = \nonumber \\
&&\langle \phi^*(k,t) \phi^*(-k,t) \phi(-k,t) \phi(k,t) \rangle \\
&& -\frac{1}{2} \langle \phi^*(k,t) \phi(k,t) \rangle -\frac{1}{2} \langle \phi^*(-k,t) \phi(-k,t) \rangle + \frac{1}{4}.  \nonumber
\end{eqnarray}

\subsection{Numerical results}

\begin{figure}
	\includegraphics[scale=0.2, angle=0]{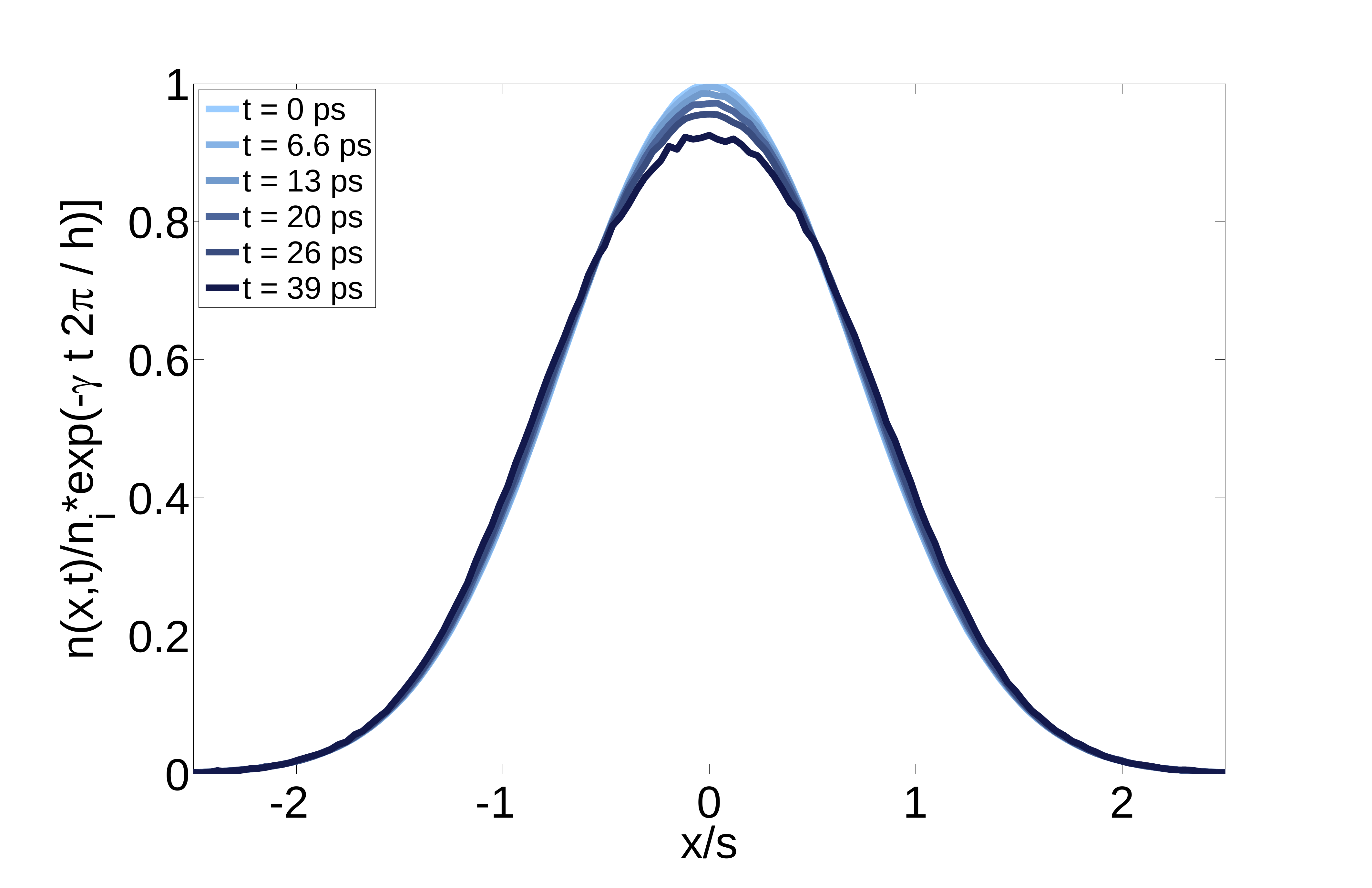}
	\caption{Density calculated with Monte Carlo simulations, initial Gaussian profile given by $\exp(-x^2/s^2)$, with $s=100\, \mu$m. We have chosen $g=0.01 \, \mu {\rm m \, meV}, \; \gamma=0.05 \, {\rm  meV}, \hbar=1, \; m=1$, $g n_i/\gamma=10$.}
	\label{fig:densitygaus}
\end{figure}

The Monte Carlo simulation was applied to study systems with an initial Gaussian density distribution, given by $n_i \exp( - x^2/ s^2)$. For the results presented here, the values $s=100\, \mu$m and $s=20\, \mu$m have been chosen and the initial central density $n_i = 50 \, \mu\textrm{m}^{-1}$. For this density, the Bogoliubov approximation and truncated Wigner approximation were still valid for the homogeneous system. For $s=100 \mu m$, the behaviour in real space can be well understood. The density at the centre of the Gaussian decreases faster than the overall exponential decay, whereas the density at the sides of the distributions shows a relative increase, see Fig. \ref{fig:densitygaus}. This would be expected from the repulsive interactions. As a result, the density distribution becomes more homogeneous.  
\begin{figure}
	\includegraphics[scale=0.2, angle=0]{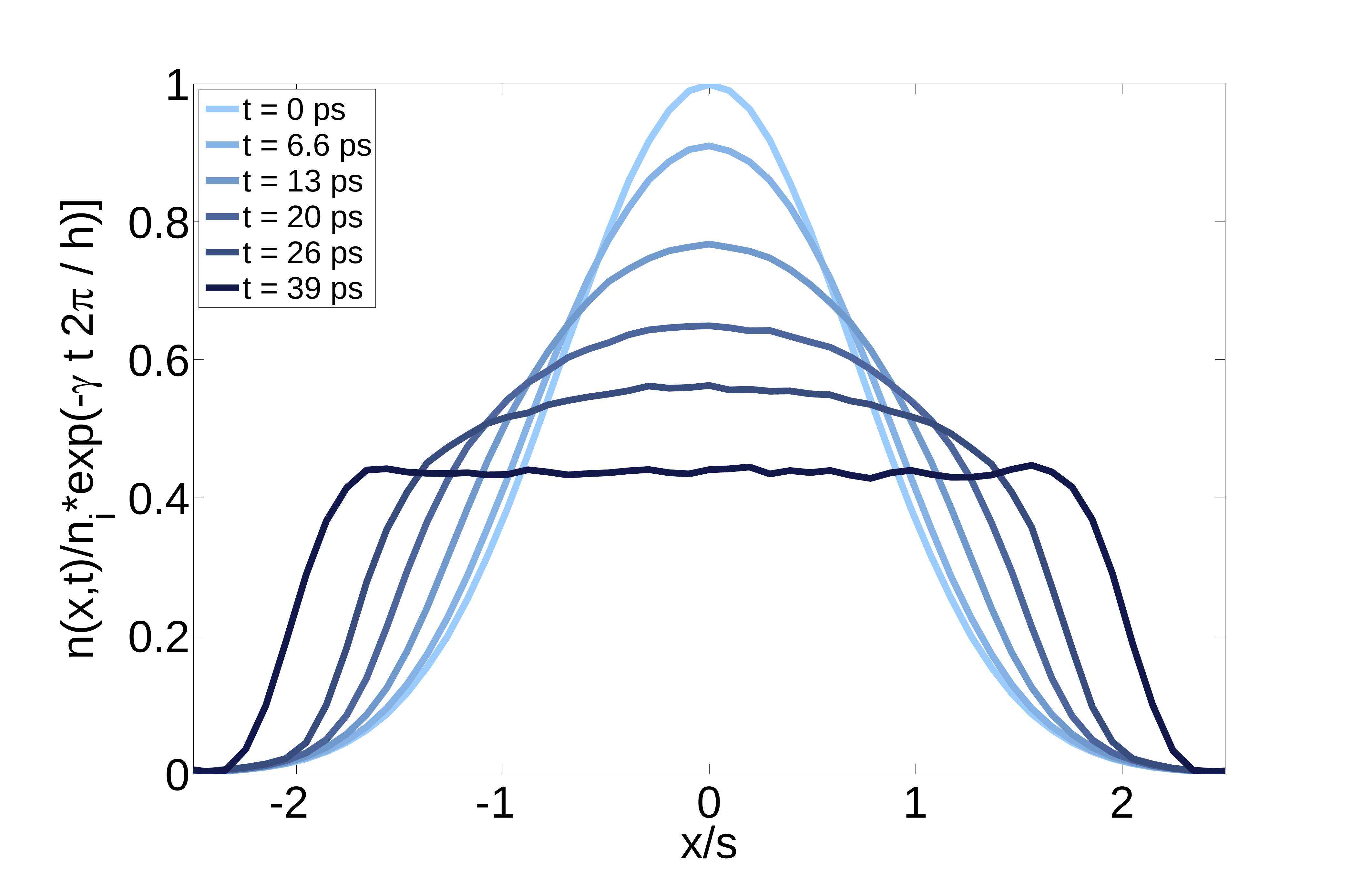}
	\caption{Density calculated with Monte Carlo simulations, initial Gaussian profile given by $\exp(-x^2/s^2)$, with $s=20\, \mu$m. We have chosen $g=0.01 \, \mu {\rm m \, meV}, \; \gamma=0.05 \, {\rm  meV},  \hbar=1, \; m=1$, $g n_i/\gamma=10$.}
	\label{fig:densitygaus20}
\end{figure}
In the case of a smaller Gaussian distribution, where $s=20\, \mu$m, this effect is even stronger. At the latest depicted time, $t=39$ps, the central region is very homogeneous, and shows a fast decay at the edges of the distribution. 

\begin{figure}
	\includegraphics[scale=0.2, angle=0]{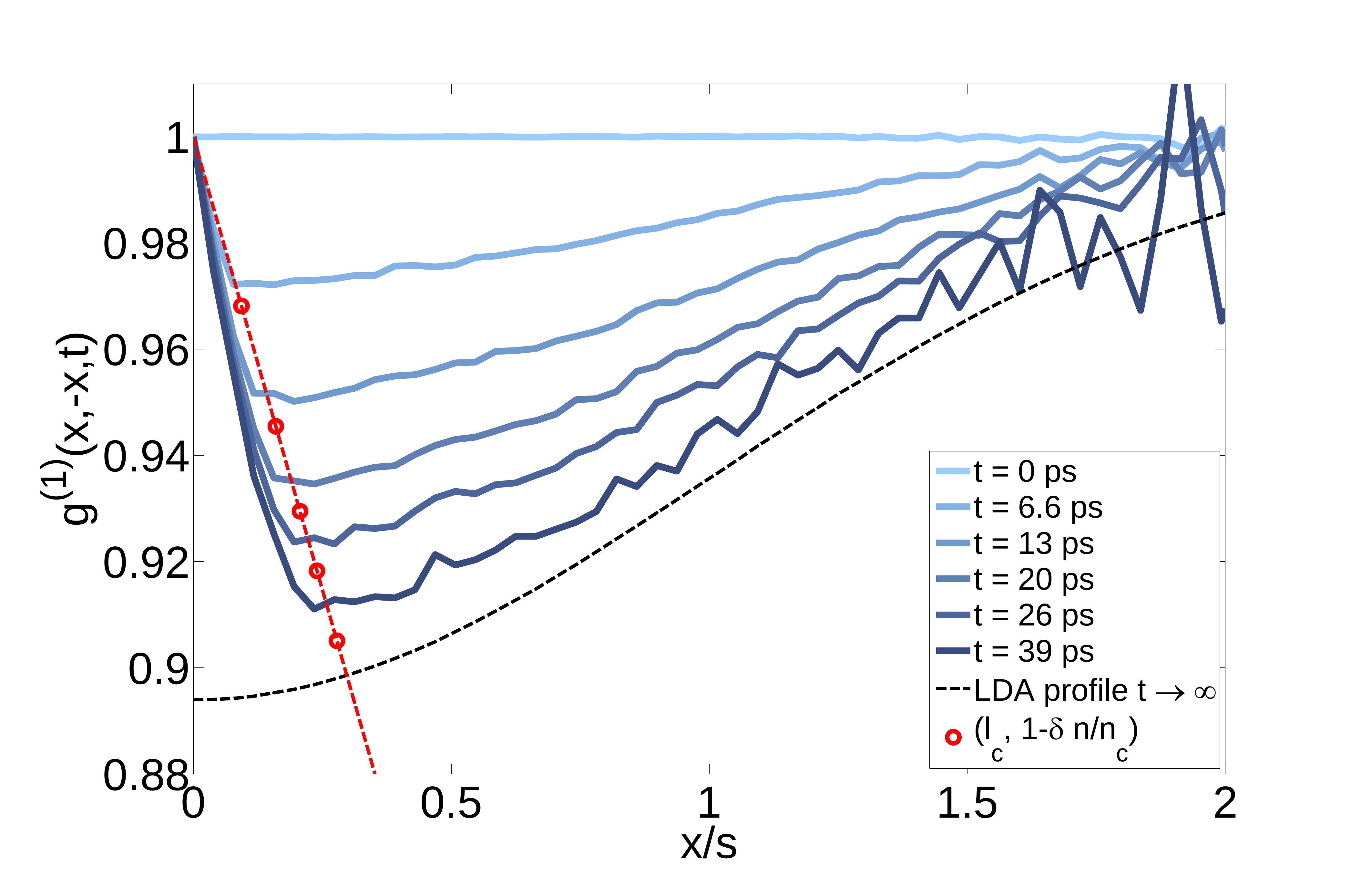}
	\caption{First order spatial coherence $g^{(1)}(x,-x,t)$ for an initial Gaussian distribution ($\exp(-x^2/s^2)$) with width $s=100\, \mu$m, calculated with Monte Carlo simulations. $\ell_c$ is given in \eqref{eq:cohlength}, and $\delta n \ n$ in \eqref{eq:dnscale}. Same parameters as Fig. \ref{fig:densitygaus}.}
	\label{fig:g1gausQMC}
\end{figure}

\begin{figure}
	\includegraphics[scale=0.2, angle=0]{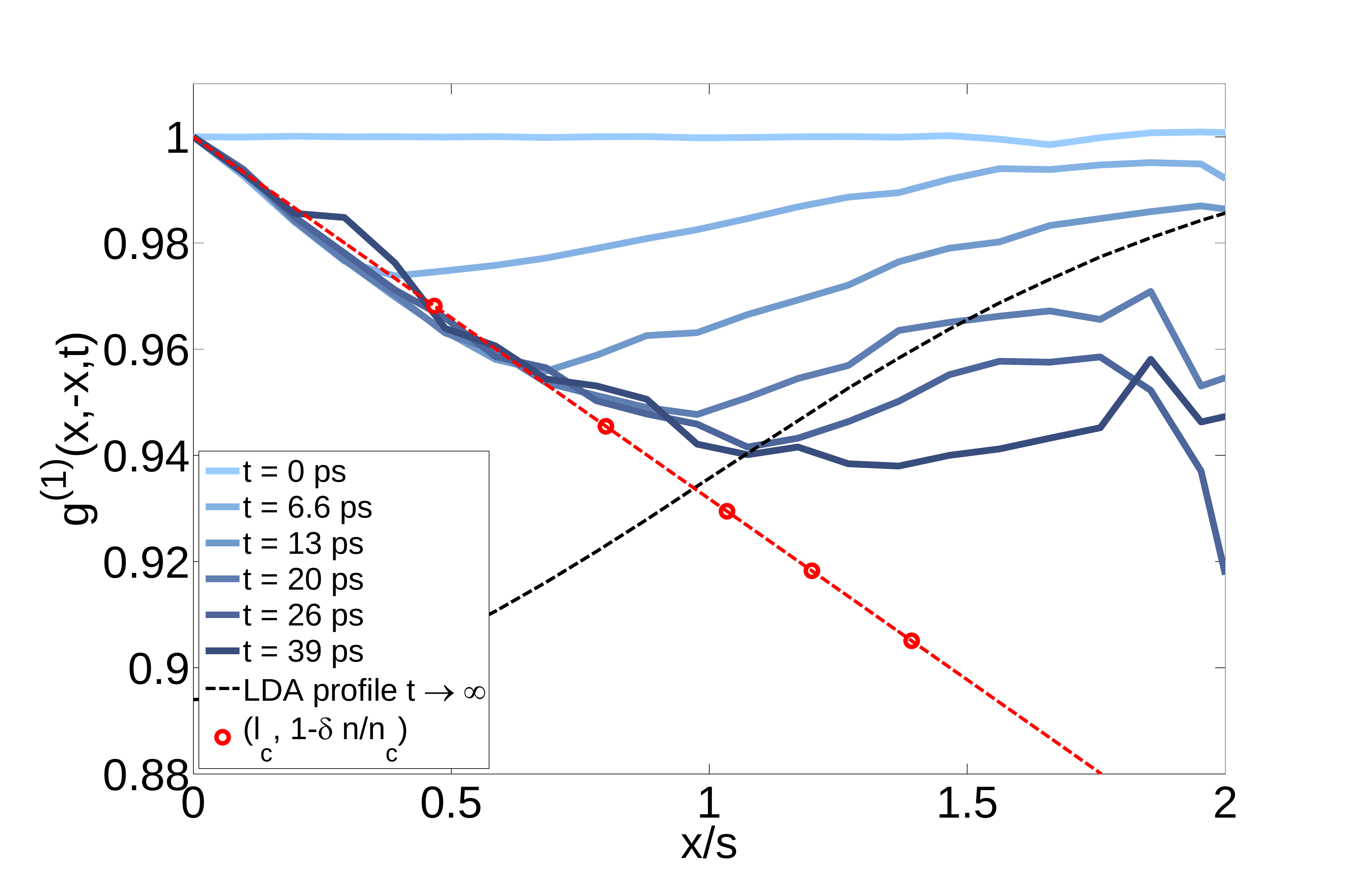}
	\caption{First order spatial coherence $g^{(1)}(x,-x,t)$ for an initial Gaussian distribution ($\exp(-x^2/s^2)$) with width $s=20\, \mu$m, calculated with Monte Carlo simulations. Same parameters as Fig. \ref{fig:densitygaus20}.}
	\label{fig:g1gausQMC20}
\end{figure}

The first order coherence for the $s=100\, \mu$m system can be described by the results obtained from the homogeneous system \cite{koghee14}. The coherence length and the maximal depletion are close to that of the homogeneous system with the density that is equal to the maximal density of the Gaussian distribution $n_i$. Since there is still a linear relation between the coherence length and the depletion, a simple correction to the numerical constants would give an even better description of these quantities. The overall shape of $g^{(1)}(x,-x,t)$ is a direct result from the Gaussian shape of the density. For the homogeneous system, a formula for the depletion at very large times as a function of the blueshift $g n_c(t)$ was derived: $\delta n/n \approx 0.77 \;  g/(\xi \gamma)$. In the inhomogeneous system the healing length $\xi=\hbar/\sqrt{m g n_c(0)}$ becomes position dependent. Consequently, the final depletion will also depend on the position. When $n_c(0)$ is simply replaced by the initial Gaussian distribution, $n_i \exp(-x^2/ s^2)$, the black dashed lines from Figs. \ref{fig:g1gausQMC} and \ref{fig:g1gausQMC20} are found. For the wider Gaussian, with $s=100\, \mu$m, this describes the behaviour of the first order coherence very well. Therefore, the phenomenon that the coherence goes back to one for large distances is due to the small density which leads to a smaller depletion. However, for the smaller Gaussian, where $s=20\, \mu$m, the local density approximation is no longer valid. Nevertheless, the initial decay of the first order coherence, until $x \approx 20 \, \mu$m is still described by the linear relation found in the homogeneous case.

\begin{figure}
	\includegraphics[scale=0.2, angle=0]{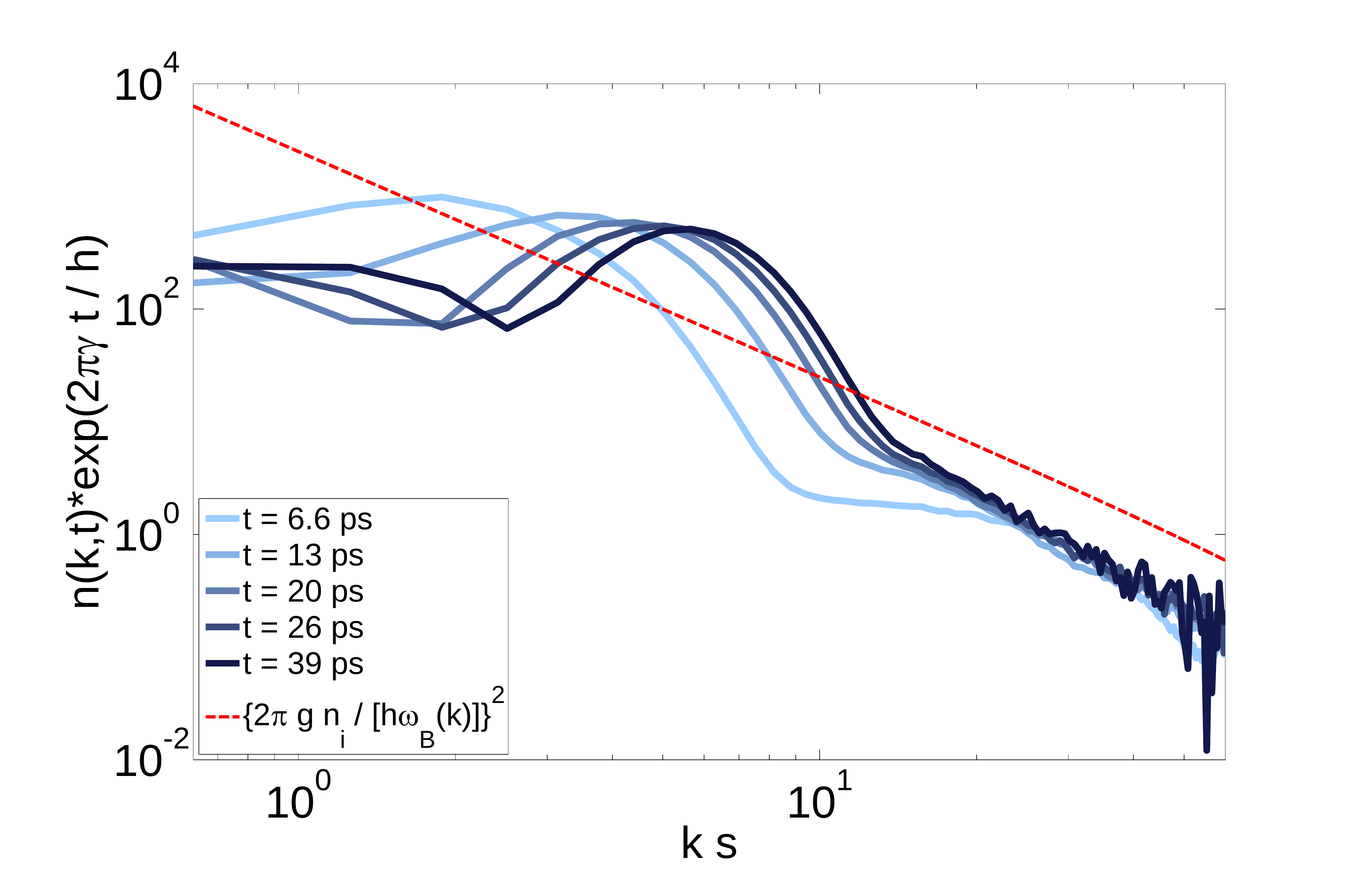}
	\caption{Momentum distribution for an initial Gaussian distribution ($\exp(-x^2/s^2)$) with width $s=100\, \mu$m, calculated with Monte Carlo simulations. Same parameters as Fig. \ref{fig:densitygaus}.}
	\label{fig:momdisgausQMC}
\end{figure}

\begin{figure}
	\includegraphics[scale=0.2, angle=0]{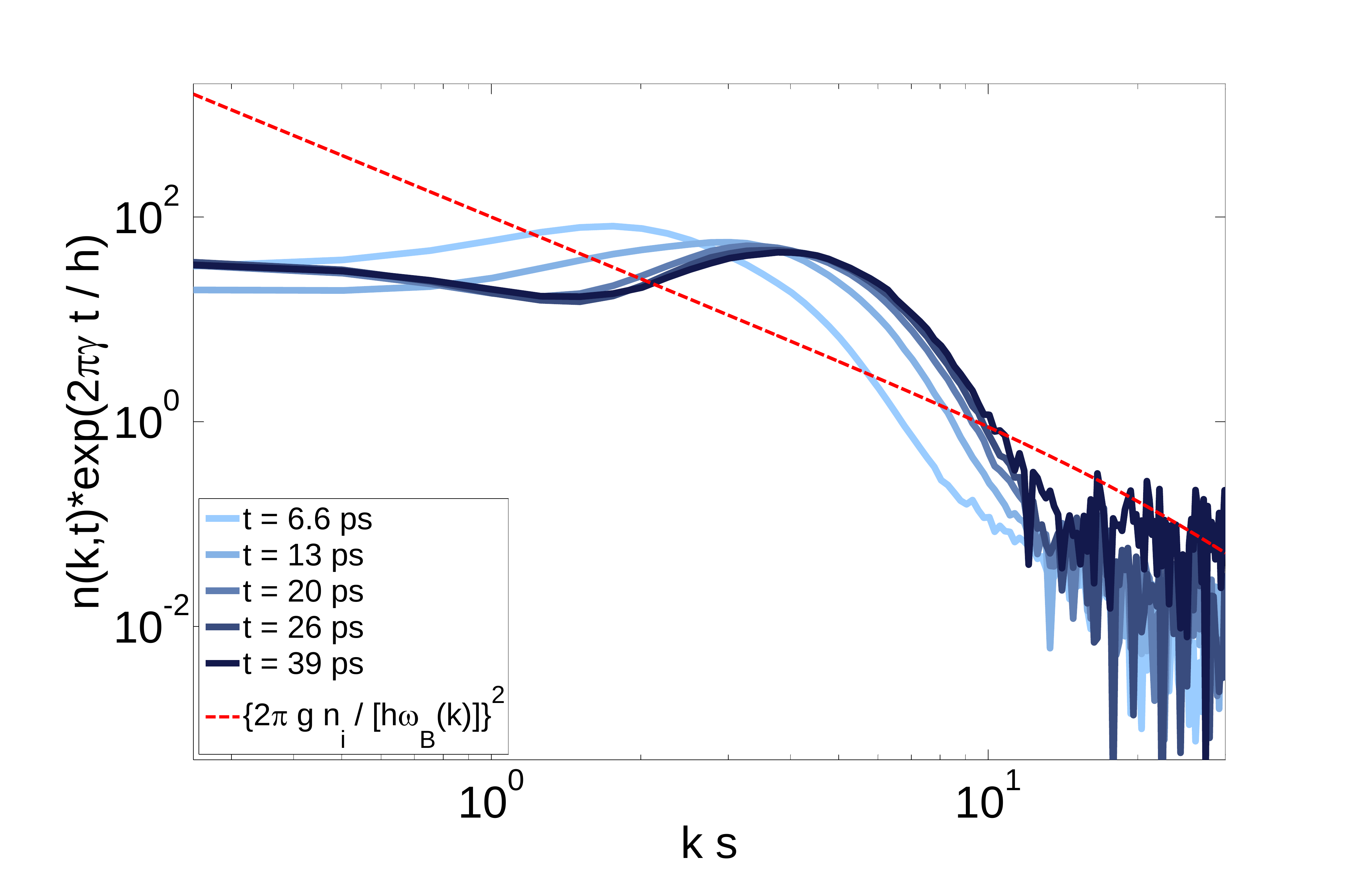}
	\caption{Momentum distribution for an initial Gaussian distribution ($\exp(-x^2/s^2)$) with width $s=20\, \mu$m, calculated with Monte Carlo simulations. Same parameters as Fig. \ref{fig:densitygaus20}. }
	\label{fig:momdisgausQMC20}
\end{figure}

In momentum space, the system with a Gaussian density is very different from a homogeneous system. This is in accordance with expectations, since for an initial Gaussian distribution, many momentum states are occupied from the start. Moreover, the inhomogeneous density profile leads to an expulsion of the polaritons away from the region where the were created. This acceleration corresponds to a shift of the momentum distribution.
 Hence, momentum conservation in the interactions no longer results in pair production of polaritons with opposite momentum. In Figs. \ref{fig:momdisgausQMC} and \ref{fig:momdisgausQMC20} we see an increase of the particle number with respect to the homogeneous case at small yet finite momentum, which indicates that the distribution is expanding, which is indeed what was seen in the evolution of the density. The wider and the smaller distribution both display very similar behavior, where the smaller density distribution has a peak in the momentum distribution at larger momenta, as compared to the wider density distribution. At larger momenta, the momentum distribution follows that of the homogeneous system more closely for the $s=100\, \mu$m case. Nevertheless, the particle number at these momenta is very small. 

\begin{figure}
	\includegraphics[scale=0.2, angle=0]{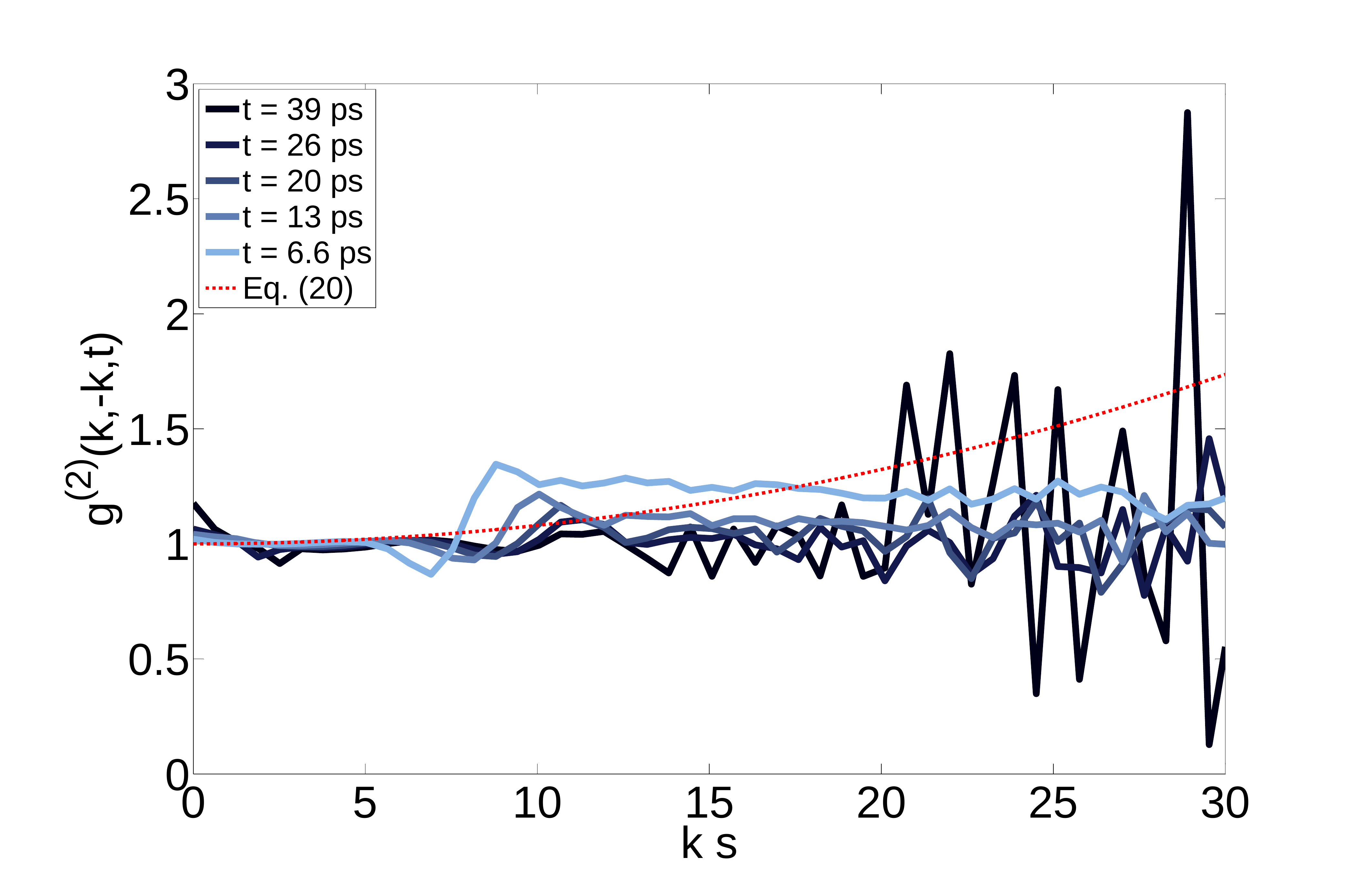}
	\caption{Second order coherence in momentum space for initial Gaussian distribution ($\exp(-x^2/s^2)$) with width $s=100\, \mu$m, calculated with Monte Carlo simulations. Same parameters as Fig. \ref{fig:densitygaus}. }
	\label{fig:g2gausQMC}
\end{figure}

\begin{figure}
	\includegraphics[scale=0.2, angle=0]{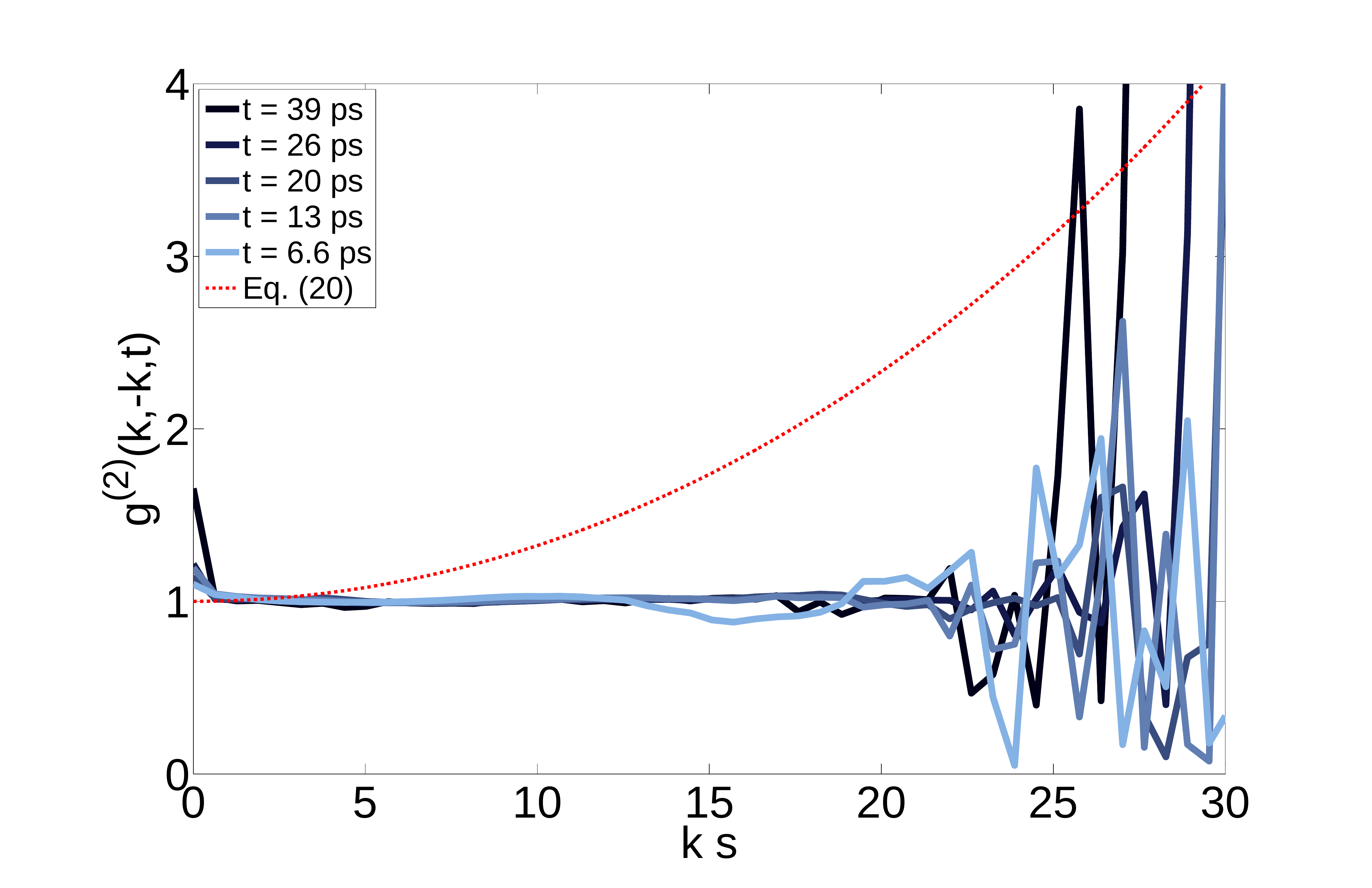}
	\caption{Second order coherence in momentum space for initial Gaussian distribution ($\exp(-x^2/s^2)$) with width $s=20\, \mu$m, calculated with Monte Carlo simulations. Same parameters as Fig. \ref{fig:densitygaus20}. }
	\label{fig:g2gausQMC20}
\end{figure}

The second order coherence, depicted in Figs. \ref{fig:g2gausQMC} and \ref{fig:g2gausQMC20}, is even more different from the homogeneous case. For small momenta, the second order coherence is close to one, which is the value corresponds to a coherent system. We see that the second order coherence remains close to one, even for momenta at which the momentum distribution seems to resemble the homogeneous case for the wider density distribution. This suggests that even for these momenta, most particles are part of the condensate. For the system with $s=20\, \mu$m,  the second order coherence, is either very close to one, or it contains too much noise for a good description. 

\section{Conclusions \label{sec:concl}}
We have studied a quantum quench consisting of a sudden injection of polaritons in a microcavity. Both a homogeneous and a Gaussian initial density distribution have been examined. The homogeneous case has been related to the dynamical Casimir effect previously. Where the correlation functions in the homogeneous case could still be calculated analytically, we performed truncated Wigner Monte Carlo simulations for the Gaussian excitation pulse. The first order spatial coherence is well approximated by the local density approximation for a sufficiently wide pulse. The second order coherence in momentum space evidences the production of excitations in pairs. For a system with an initial Gaussian density distribution, multiple momentum states are significantly occupied from the start. Here, the second order coherence indicates that many of these particles are coherent, so that evidence of quantum correlations is highly suppressed.

\end{document}